\documentclass[runningheads]{llncs}
\usepackage{graphicx}
\usepackage{amsmath}
\usepackage{amssymb}
\usepackage{hyperref}
\usepackage{color}

\usepackage[ruled,vlined,linesnumbered]{algorithm2e}
\usepackage{algpseudocode}
\usepackage{booktabs}
\usepackage{todonotes}
\usepackage{comment}
\usepackage[group-separator={,}]{siunitx}
\usepackage[normalem]{ulem}
\usepackage{mathtools}

\newcommand{\expSym}{\mathbb{E}}
\newcommand{\E}[1]{\ensuremath{\expSym\left(#1\right)}}

\newcommand{\T}{\top}

\newcommand{\Cov}[2]{\ensuremath{\mathrm{Cov}(#1, #2)}}
\newcommand{\vw}[1]{}
\newcommand{\MB}[1]{}

\newcommand{\new}[1]{#1}

\renewcommand{\inst}[1]{\textsuperscript{#1}}

\DeclareMathOperator*{\argmax}{arg\,max}

\newtheorem{model}{Model}

\begin{document}
\title{Variance Reduction\\  in Stochastic Reaction Networks\\ using Control Variates}
\titlerunning{Variance Reduction\\  in SRNs \\ using Control Variates}
\author{Michael Backenk\"ohler\inst{1,2}, Luca Bortolussi\inst{3,1}, Verena Wolf\inst{1}}
\authorrunning{M.\ Backenk\"ohler et al.}
\institute{\inst{1}Saarland University, Germany,\\ \inst{2} Saarbr\"ucken Graduate School of Computer Science,\\
\inst{3}University of Trieste, Italy}
\maketitle
\begin{abstract}
\new{Monte Carlo estimation in plays a crucial role in stochastic 
reaction networks.
However,  reducing the statistical uncertainty of the corresponding estimators requires sampling a large number of trajectories.
We propose control variates based on the statistical moments of the process to reduce the estimators' variances.
We develop an algorithm that selects an efficient subset of infinitely many control variates.
To this end, the algorithm uses resampling and a redundancy-aware greedy selection.
We demonstrate the efficiency of our approach in several case studies.
}
\keywords{Chemical Reaction Network \and Stochastic Simulation Algorithm
\and Moment Equations \and Control Variates \and Variance Reduction \and Monte Carlo}
\end{abstract}
\section{Introduction}
Stochastic reaction networks that are used to describe cellular processes are often subject to inherent stochasticity.
The dynamics of gene expression, for instance, is influenced by 
single random events (e.g.\ transcription factor binding) and 
hence, models that take this randomness into account must monitor
discrete molecular counts and reaction events that change these counts.
Discrete-state continuous-time Markov chains have successfully  been
used to describe  networks of chemical reactions
over time that correspond to the basic events of such processes. 
The time-evolution of the corresponding probability distribution is 
given by the chemical master equation,
\new{which is a system of differential equations with one equation for each possible molecular count. However, its  } numerical solution is
extremely challenging because of the enormous size of the underlying
state-space. 

\new{In contrast,} analysis approaches based on sampling, such as the Stochastic Simulation Algorithm (SSA) \cite{gillespie77}, can  be applied
independent of the size of the model's state-space. 
However, statistical approaches are costly since a large number
of simulation runs is necessary to reduce the statistical 
inaccuracy of estimators. This problem is particularly severe
if reactions occur on multiple time scales or if the event of interest is rare.
A particularly popular technique to speed up simulations is $\tau$-leaping which applies
multiple reactions in one step of the simulation.
However, such multi-step simulations rely on certain assumptions about
the number of reactions in a certain time interval. These assumptions
are typically only approximately fulfilled and therefore introduce  approximation
errors on top of the statistical uncertainty of the considered point estimators.

\new{
Variance reduction techniques are an alternative to approaches that decrease the computational costs of each SSA run. By reducing the variance of the estimators, these methods need fewer runs to achieve high statistical accuracy.
}

In this work, we approach the variance reduction problem by considering 
a (infinite) set of differential equations for the evolution of the statistical moments of the molecular counts. Instead of applying a moment closure and performing numerical integration \cite{ale2013general,engblom2006computing}, we use these equations 
to derive a combination of   moment constraints. Such moment constraints have already been used for for parameter
estimation \cite{backenkohler2018moment} and for computing moment bounds using semi-definite programming \cite{dowdy2018dynamic,ghusinga2017exact}.
Here, we   interpret these constraints as random variables 
that are correlated with the estimators of interest usually given as functions of population variables.
These constraints can be used  as  (linear) control variates in order to improve the final estimate and reduce its variance \cite{lavenberg1982statistical,szechtman2003control}.
The method is easy on an intuitive level: If a control variate
is positively correlated with the function to be estimated then 
we can use the estimate of the variate to
adjust the target estimate.

The incorporation of control variates into the SSA
introduces   additional simulation costs for the calculation of the constraint values.
These values are integrals over time, which we accumulate  based on the
piece-wise constant trajectories.
This introduces a trade-off between the variance reduction that is achieved
by using control variates versus the increased simulation cost.
This trade-off is expressed as the product of the variance reduction ratio
and the cost increase ratio.

For a good trade-off, it is crucial to find an appropriate set of control variates.
Here we propose a class of constraints which is parameterized by a moment vector
and a weighting parameter, resulting in infinitely many choices.

\new{
In previous work \cite{backenkohler2019control}, we have proposed an algorithm that learns a set of
control variates through refinement of an initial set.
This initial set of control variates is based on samples of the time-weighting $\lambda$.
Each control variate is then checked for effectiveness in isolation.
Furthermore, the set is refined by considering variables pairwise to determine redundancies.

In this work, we improve on the initial selection of control variates.
This initial set  is build using a splitting approach akin to sequential Monte Carlo methods:
Over multiple rounds, new control variates are sampled based on their performance in prior iterations.
This way, we construct a set of candidate variates and select a subset using a greedy approach, which takes into account the correlation between variates.
A benefit of this algorithm is that it is less sensitive to user input.
In particular, no heuristic redundancy threshold has to be fixed, making this approach more flexible.
}

This approach applies to the Monte Carlo estimation of any quantity deal with any property that can
be expressed in terms of expected values such as probabilities
of complex path properties.
Another advantage of our technique is that increased efficiency is achieved without the price of an additional approximation error as is the case for methods based on moment approximations or multi-step simulations.

This paper is structured as follows. In Section~\ref{sec:related} we give a brief
survey of  methods and tools related to efficient stochastic simulation
and moment techniques. In Section~\ref{sec:bg} we introduce the
common stochastic semantics of stochastic reaction networks. From these
semantics we show in Section~\ref{sec:moments} how to derive
constraints on the moments of the transient distribution.
The variance reduction technique of control variates is described in
Section~\ref{sec:var_red}.
We show the design of an algorithm using moment constraints to reduce
sample variance in Section~\ref{sec:algo}.
The efficiency and other characteristics of this algorithm are evaluated
on four non-trivial case studies in Section~\ref{sec:study}.
Finally, we discuss the findings and give possibilities for further
work in Section~\ref{sec:conclusion}.

\section{Related Work}\label{sec:related}
Much research has been directed at the efficient analysis of stochastic stochastic reaction
networks.
Usually research focuses on improving efficiency by making certain approximations.

If the state-space is finite and small enough one can deal with the underlying Markov
chain directly.
But there are also  cases where the transient distribution has an infinitely large support
and one can still deal with explicit state probabilities.
To this end, one can fix a finite state-space, that should contain most of the
probability~\cite{munsky2006finite}. Refinements of the method work
dynamically and adjust the state-space according to the transient
distributions~\cite{andreychenko2011parameter,henzinger2009sliding,mateescu2010fast}.

On the other end of the spectrum there are mean-field approximations, 
which model the mean densities faithfully in the system size limit~\cite{bortolussi2013continuous}.
In between there are techniques such as moment closure \cite{singh2006lognormal}, that
not only consider the mean, but also the variance and other higher order moments.
These methods depend on ad-hoc approximations of higher order moments to
close the ODE system given by the moment equations.
Yet another class of methods approximate molecular counts continuously and approximate the dynamics in such a continuous space, e.g. the system size expansion~\cite{van1992stochastic} and the
chemical Langevin equation~\cite{gillespie2000chemical}.

While the moment closure method uses ad-hoc approximations for high order moments to
facilitate numerical integration, they can be avoided in some contexts.
For the equilibrium distribution, for example, the time-derivative of all moments is equal to zero.
This directly yields constraints that have been used for parameter estimation at
steady-state~\cite{backenkohler2018moment}
and bounding moments of the equilibrium distribution using semi-definite
programming~\cite{ghusinga2017estimating,ghusinga2017exact,kuntz2017rigorous}.
The latter technique of bounding moments has been successfully adapted in the 
context of transient analysis~\cite{dowdy2018dynamic,sakurai2017convex,sakurai2019bounding}.
We adapt the constraints proposed in these works to improve statistical estimations via stochastic
simulation (cf.\ section~\ref{sec:moments}).

While the above techniques give a deterministic output, stochastic simulation generates
single executions of the stochastic process~\cite{gillespie77}.
This necessitates accumulating large numbers of simulation runs to estimate
quantities.
This adds a significant computational burden. Consequently, some effort
has been directed at lowering this cost.
A prominent technique is $\tau$-leaping~\cite{gillespie2001approximate},
which in one step performs multiple instead of only a single reaction.
Another approach is to find approximations that are specific to the problem at hand,
such as approximations based on time-scale separations~\cite{cao2005slow,bortolussi2015efficient}.

Multilevel Monte Carlo methods have been applied in to time-inhomogeneous
SRNs~\cite{anderson2018low}. In this techniques estimates are combined
using estimates of different approximation levels.

\new{
 In the case of rare events, approaches based on importance sampling \cite{gillespie2009refining,roh2010state} and importance splitting \cite{jegourel2013importance} have been adapted to the setting of reaction networks.
Importance sampling relies on a suitable change of the underlying probability measure, which is often handcrafted for each model, continually refined using the cross-entropy method \cite{daigle2011automated,roh2011state}, or derived from Gaussian approximations of the process \cite{gillespie2019guided}.
Importance splitting decomposes the state space into level sets and estimates the rare event probability by a level-based splitting of sample paths  before reaching the set of interest. It requires to construct a model-specific level function together with the corresponding splitting thresholds.

Recently, randomized quasi-Monte Carlo (RQMC) approaches have been adapted to the application area of stochastic reaction networks \cite{beentjes2019quasi} and improved for the case of long simulation horizons with an extension to array-RQMC \cite{puchhammer2021variance}. It is based on a tau-leaping approach where time is discretized and requires a level/importance function or a costly multivariate sort.
}

\section{Stochastic Reaction Networks}\label{sec:bg}
A stochastic reaction network (SRN) describes the interactions between
a set of species $S_1,\dots, S_{n_S}$ in a well-stirred reactor.
Since we assume that all reactant molecules are spatially uniformly distributed,
we just keep track of the overall amount of each molecule.
Therefore the state-space is given by $\mathcal{S}\subseteq\mathbb{N}^{n_S}$.
These interactions are expressed a set of \emph{reactions} with a certain
inputs and outputs, given by the vectors $v_j^{-}$ and $v_j^{+}$ for
reaction $j=1,\dots,n_R$, respectively. Such reactions are denoted as
\begin{equation}\label{eq:reaction}
    \sum_{i=1}^{n_S} v_{ji}^{-} S_i
    \xrightarrow{c_j}
    \sum_{i=1}^{n_S} v_{ji}^{+} S_i\,.
\end{equation}
The reaction rate constant $c_j>0$ gives us information on the propensity
of the reaction.
If just a constant is given, \emph{mass-action} propensities are assumed.
In a stochastic setting for some state $x\in\mathcal{S}$ these are
\begin{equation}\label{eq:stoch_mass_action}
    \alpha_j(x)=c_j\prod_{i=1}^{n_S}\binom{x_i}{v_{ji}^{-}}\,.
\end{equation}
The system's behavior is described by a stochastic
process $\{X_t\}_{t\geq 0}$.
The propensity function gives the infinitesimal probability
of a reaction occurring, given a state $x$. That is, for a
small time step $\delta t >0$
\begin{equation}\label{eq:reaction_prob}
    \Pr(X_{t+\delta t}=x+v_j\mid X_t=x)
    =\alpha_j(x)\delta t + o(\delta t)\,.
\end{equation}
This induces a corresponding continuous-time
Markov chain (CTMC) on $\mathcal{S}$ with generator matrix\footnote{Assuming a fixed enumeration of the state space.}
\begin{equation}\label{eq:cme_generator}
    Q_{x,y} = \begin{cases}
        \sum_{j:x+v_j = y}\alpha_j(x)\,,&\text{if}\;x\neq y\\
        -\sum_{j=1}^{n_R} \alpha_j(x)\,, &\text{otherwise.}
    \end{cases}
\end{equation}
Accordingly, the time-evolution of the process' distribution,
given an initial distribution $\pi_0$, is given by the
Kolmogorov forward equation, i.e.\ $\frac{d\pi_t}{dt}=Q\pi_t$, where $\pi_t(x)=\Pr(X_t=x)$.
For a single state, it is commonly referred to 
as the \emph{chemical master equation} (CME)
\begin{equation}\label{eq:cme}
    \frac{d}{d t} \pi_t(x) =
    \sum_{j=1}^{n_R}\left(
        \alpha_j(x-v_j)\pi_t(x-v_j) - \alpha_j(x)\pi_t(x)
    \right)\,.
\end{equation}
A direct solution of \eqref{eq:cme} is usually not possible.
If the state-space with non-negligible probability is suitably small, a state space
truncation could be performed. That is, \eqref{eq:cme} is integrated on a possibly time-dependent subset
$\hat{\mathcal{S}}_t\subseteq\mathcal{S}$ \cite{henzinger2009sliding,munsky2006finite,spieler2014numerical}.
Instead of directly analyzing \eqref{eq:cme}, one often resorts to simulating trajectories.
A trajectory $\tau=x_0t_1x_1t_1\dots t_n x_n$ over the interval $[0,T]$ is a sequence of states $x_i$
and corresponding
jump times $t_i$, $i=1,\dots,n$ and $t_n=T$.
We can sample trajectories of $X$ by using stochastic simulation~\cite{gillespie77}.

Consider the birth-death model below as an example.
\begin{model}[Birth-death process]\label{model:bd}
    A single species $\mathsf{A}$ has a constant
    production and a decay that is linear in the current amount of molecules.
    Therefore the model consists of two mass-action reactions
    $$\varnothing \xrightarrow{\gamma} \mathsf{A}\,,\quad
    \mathsf{A}\xrightarrow{\delta}\varnothing\,,$$
    where $\varnothing$ denotes   no reactant or no product,
    respectively.
\end{model}
For Model~\ref{model:bd} the change of probability mass in a single state $x>0$ is described by expanding
\eqref{eq:cme} and
$$\frac{d}{dt}\pi_t(x)=\gamma \pi_t(x-1) + \delta \pi_t(x+1) - (\gamma + \delta)\pi_t(x)\,.$$
We can generate trajectories of this model by choosing either reaction, with a probability that is
proportional to its rate given the current state $x_i$.
The jump time $t_i- t_{i+1}$ is determined by sampling from an exponential distribution with rate $\gamma+x_i\delta$.

\section{Moment Constraints}\label{sec:moments}
The time-evolution of $\E{f(X_t)}$ for some function $f$ can be directly derived from \eqref{eq:cme} by 
computing the sum
$\sum_{x\in\mathcal{S}}f(x)\frac{d}{dt}\pi_t(x)$, which yields 
\begin{equation}\label{eq:mom_ode}
    \frac{d}{dt}\E{f(X_t)} = \sum_{j=1}^{n_R}\E{\left(f({X_t + v_j}) - f(X_t)\right)\alpha_j(X_t)}\,.
\end{equation}
While many choices of $f$ are possible, for this work we will restrict ourselves to
monomial functions $f(x)=x^m$, $m\in\mathbb{N}^{n_S}$
i.e.\ the \emph{non-central moments} of the process.
The \emph{order} $\lvert m\rvert$ of a moment $\E{X^m}$ is the sum over the exponents,
i.e.\ $\lvert m\rvert =\sum_im_i$.
The integration of \eqref{eq:mom_ode} with such functions $f$ is well-known in the context of
moment approximations of SRN models.
For most models the arising ODE system is infinitely large, because the time-derivative of
low order moments usually depends on the values of higher order moments.
To close this system, \emph{moment closures}, i.e.\ ad-hoc approximations of higher order moments
are applied \cite{schnoerr2015comparison}.
The main drawback of this kind of analysis is that it is not known whether the chosen closure gives
an accurate approximation for the case at hand.
Here, such approximations are not necessary, since we will apply the moment dynamics in the context
of stochastic sampling instead of trying to integrate \eqref{eq:mom_ode}.

Apart from integration strategies,
setting \eqref{eq:mom_ode} to zero has been used as a constraint for parameter estimation at steady-state
\cite{backenkohler2018moment} and bounding moments at steady-state~\cite{dowdy2018bounds,ghusinga2017exact,kuntz2017rigorous}.
The extension of the latter has recently lead to the adaption of these constraints
to a transient setting~\cite{dowdy2018dynamic,sakurai2019bounding}.
These two transient constraint variants are analogously derived by multiplying \eqref{eq:mom_ode}
by a time-dependent, differentiable weighting function $w(t)$ and integrating:

Multiplying with $w(t)$ and integrating on $[t_0, T]$ yields~\cite{dowdy2018dynamic,sakurai2019bounding}
\begin{equation}\label{eq:poly_con}
\begin{split}
        & w(T)\E{f(X_{T})}
        - w(t_0)\E{f(X_{t_0})}
        - \int_{t_0}^{T}\frac{dw(t)}{dt}\E{f(X_t)}\,dt\\
        =&\sum_{j=1}^{n_R}\int_{t_0}^{T}w(t)
        \E{\left(f{(X_t + v_j)} - f(X_t)\right)\alpha_j(X_t)}\,dt
        \end{split}
\end{equation}

In the context of computing moment bounds via semi-definite programming
the choices $w(t)=t^s$~\cite{sakurai2019bounding} and $w(t)=e^{\lambda(T - t)}$~\cite{dowdy2018dynamic}
have been proposed.
While both choices proved to be effective in different case studies, relying solely on the latter choice,
i.e.\ $w(t)=e^{\lambda(T - t)}$ was sufficient. \new{We can further forgo the time inversion such that
$w(t)=e^{\lambda t}$.}

By expanding the rate functions and $f$ in \eqref{eq:poly_con} and substituting the
exponential weight function we can re-write \eqref{eq:poly_con} as
\begin{equation}\label{eq:simpl_exp}
        0 =\,
         e^{\lambda T}\E{f(X_{T})}
        - \E{f(X_{t_0})}
        + \sum_{k}c_k\int_{t_0}^{T}e^{\lambda t}\E{X_t^{m_k}}\,dt
\end{equation}
with coefficients $c_k$ and vectors $m_k$ defined accordingly.
Assuming the moments remain finite on $[0,T]$, we can define the random variable
\begin{equation}\label{eq:z}
        Z =\,
         e^{\lambda T}f(X_{T})
        - f(X_{t_0})
        + \sum_{k}c_k\int_{t_0}^{T}e^{\lambda t}X_t^{m_k}\,dt
\end{equation}
with $\E{Z}=0$.

Note, that a realization of $Z$ depends on the whole trajectory $\tau=x_0t_1 x_1 t_1 \dots\allowbreak t_n x_n$ over $[t_0,T]$.
Thus, for the integral terms
in \eqref{eq:z} we have to compute sums
\begin{equation}\label{eq:dis_int}
    \frac{1}{\lambda}\sum_{i=1}^n\left(e^{\lambda t_{i+1}}
    - e^{\lambda t_i}\right)x_i^{m_k}\,,
\end{equation}
over a given trajectory.
This accumulation is best done during the simulation to avoid storing the whole trajectory.
\new{Algorithm~\ref{alg:ssa_lcv} specifies the stochastic simulation of system trajectories while computing the values of integrals \eqref{eq:dis_int} alongside \vw{with}  the trajectory itself.}
\begin{algorithm}
    \SetKwInOut{Input}{input}
    \SetKwInOut{Output}{output}
    \Input{$\pi_0, T, P, n$}
    \Output{trajectory $\tau$}
    initialize accumulator map $A$
    \For{$i=1,\dots, n$}{
        $\tau \leftarrow$ empty list, $s\leftarrow$ sample from $\pi_0$, $t\leftarrow 0$\;
        \While{$t<T$}{
        $\tau\leftarrow \text{append}(\tau, (s, t))$\;
            $k\leftarrow$ sample reaction $i$ with probability $\propto\alpha_i(s)$\;
            $\delta\sim \text{Exp}\left(\sum_i \alpha_i(s)\right)$\;
            \For{$(m,\lambda)\in \mathit{keys}(A)$}{
                $A[(m,\lambda)]\leftarrow A[(m,\lambda)] + \frac{1}{\lambda}\left(e^{ \lambda(t + \delta)} - e^{ \lambda t }\right)x^m$\;
            }
            $s\leftarrow s + v_k$\;
            $t \leftarrow t + \delta$\;
        }
        update means $\hat{V}$, $\hat{Z}$ and covariances $\hat\Sigma$ using $A$\;
        \For{$(m,\lambda)\in \mathit{keys}(A)$}{$A[(m,\lambda)] \leftarrow 0$}
    }
    \textbf{return} $(\hat{\Sigma},\hat{V}, \hat{Z})$\;
    \caption{\label{alg:ssa_lcv}SSA with accumulator updates}
\end{algorithm}
The cost of a simulation using this algorithm is more expensive.
For the method to be efficient, the variance reduction (Section~\ref{sec:var_red}) needs
to overcompensate for this increased cost of a simulation run.

For Model~\ref{model:bd} the moment equation for $f(x)=x$ becomes
$$\frac{d}{dt}\E{X_t}=\gamma - \delta\E{X_t}\,.$$
The corresponding constraint \eqref{eq:simpl_exp} with $\lambda=0$ gives
$$0=\E{X_T} - \E{X_0} - \gamma T + \delta \int_0^{T} \E{X_t}\,dt\,.$$
In this instance the constraint  leads to an explicit function of
the moment over time. If  $X_0=0$ w.p.\ 1, then \eqref{eq:simpl_exp} becomes
\begin{equation}\label{eq:bd_constraint}
\E{X_T} = \frac{\gamma}{\delta} \left(1 - e^{-\delta T}\right)
\end{equation}
when choosing $\lambda=\delta$.

\section{Control Variates}\label{sec:var_red}
Now, we are interested in the estimation of some quantity $\E{V}$
by stochastic simulation.
Let $V_1,\dots,V_n$ be independent samples of $V$.
Then the sample mean $\hat{V}_n
=\frac{1}{n}\sum_{i=1}^n V_k$ is an estimate of $\E{V}$.
By the central limit theorem
\[
\sqrt{n}\hat{V}_n\xrightarrow{d}N(\E{V},\sigma_V^2)\,.
\]
Now suppose, we know of a random variable $Z$ with $0=\E{Z}$.
The variable $Z$ is called a \emph{control variate}.
If a control variate $Z$ is correlated with $V$, we can
use it to
reduce the variance of $\hat{V}_n$~\cite{glasserman2005large,nelson1990control,szechtman2003control,wilson1984variance}.
For example, consider we are running a set of simulations and consider a single
constraint.
If the estimated value of this constraint is larger than zero and we estimate a positive correlation
between the constraint $Z$ and $V$, we would, intuitively, like to {decrease} our
estimate $\hat{V}_n$ accordingly.
This results in an estimation of the mean of the random variable $$Y_{\beta}= V-\beta Z$$ instead of $V$.
The variance
$$\sigma_{Y_{\beta}}^2 = \sigma_V^2-2\beta \Cov{V}{Z} + \beta^2\sigma_Z^2\,.$$
The optimal choice $\beta$ can be computed by  considering the minimum of $\sigma_{Y_\beta}^2$. Then
$$\beta^{*}={\Cov{V}{Z}}/{\sigma_Z^2}\,.$$
Therefore $\sigma_{Y_{\beta^{*}}}=\sigma_Z^2(1 - \rho_{VZ}^2)$,
where $\rho_{VZ}$ is the correlation of $Z$ and $V$.

If we have multiple control variates, we can proceed in a similar fashion.
Now, let ${Z}$ denote a vector of $d$ control variates and let
\[
\Sigma=
\begin{bmatrix}
\Sigma_{ Z} & \Sigma_{V Z}\\
\Sigma_{ Z V} & \sigma_V^2
\end{bmatrix}
\]
be the covariance matrix of $({Z},V)$.
As above, we estimate the mean of
$
    {Y}_{\beta}=V -{\beta}^{\T}{Z}\,.
$
The ideal choice of $\beta$ is the result of an ordinary least squares regression between $V$
and $Z_i$, $i=1,\dots,n$.
Specifically, $\beta^{*}={\Sigma_{ Z}}^{-1}{\Sigma}_{ Z V}$.
Then, asymptotically
the variance of this estimator is~\cite{szechtman2003control},
\begin{equation}\label{eq:lcv_asym}
    {\sigma_{\hat Y_{\beta^*}}^2} = (1 - R_{ Z V}^2){\sigma_{\hat V}^2}\,, \quad
    R_{ Z V}^2=\Sigma_{ Z V}\Sigma_{ Z}^{-1}\Sigma_{ Z V} / \sigma_V^2\,.
\end{equation}
\new{This is commonly known as the fraction of variance unexplained \cite{freedman2009statistical}.}
In practice, however, $\beta^*$ is unknown and needs to be replaced by
an estimate $\hat{\beta}$.
This leads to an increase in the estimator's variance.
Under the assumption of $Z$ and $V$ having a multivariate normal
distribution~\cite{cheng1978analysis,lavenberg1982statistical}, the variance of the estimator is
$\hat{Y}_{\hat{\beta}}=\hat{V}-\hat{\beta}^{\top}\hat{ Z}$
\begin{equation}\label{eq:lcv_norm_varred}
    {\sigma_{\hat{Y}_{\hat{\beta}}}^2} = \frac{n - 2}{n - 2 - d}(1 - R_{ ZV}^2){\sigma_{\hat V}^2}\,.
\end{equation}

Clearly, a control  variate is ``good'' if it is highly correlated with $V$.
The constraint in \eqref{eq:bd_constraint} is an example of the extreme case.
When we use this constraint as a control variate
for the estimation of the mean at some time point $t$, it has a correlation of $\pm1$
since it describes the mean at that time precisely.
Therefore the variance is reduced to zero.
We thus aim to pick control  variates that are highly correlated with $V$.

Consider, for example, the above case of the birth-death process.
If we choose \eqref{eq:bd_constraint} as a constraint, it would always yield
the exact difference of the exact mean to the sample mean and therefore have a 
perfect correlation. Clearly, $\hat\beta$ reduces to 1 and $\hat Y_1 = \E{X_t}$.
\begin{figure}
    \centering
    \includegraphics[scale=.65]{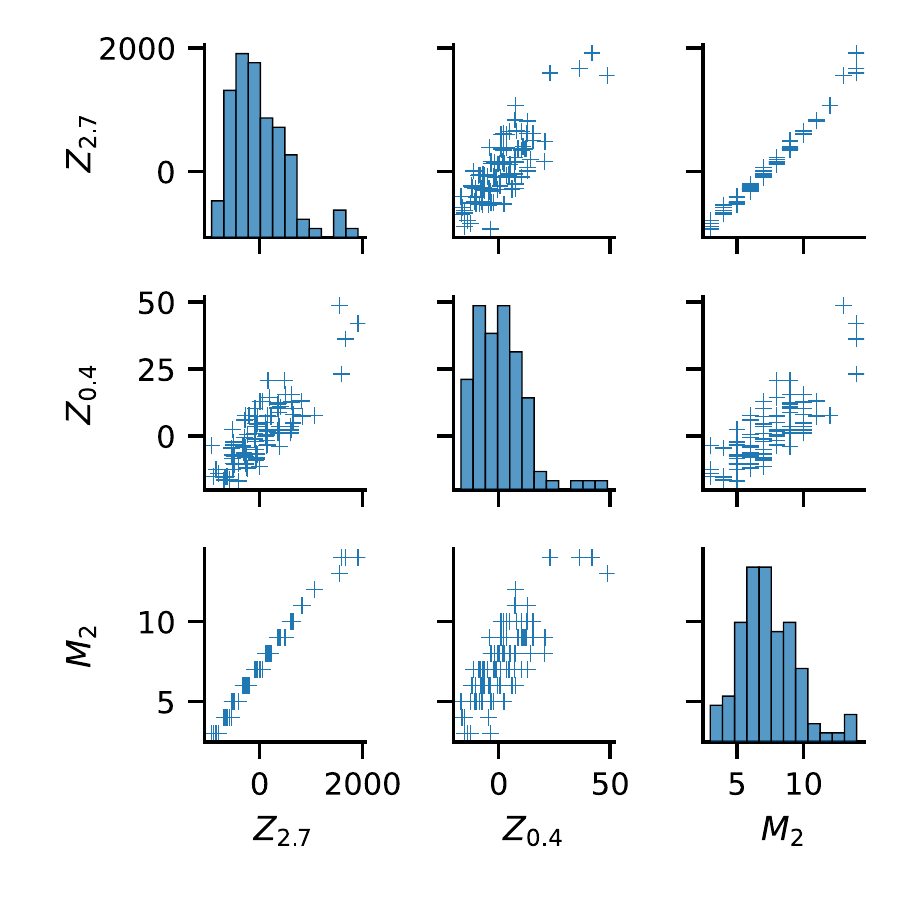}
    \hspace{2em}
    \includegraphics[scale=.65]{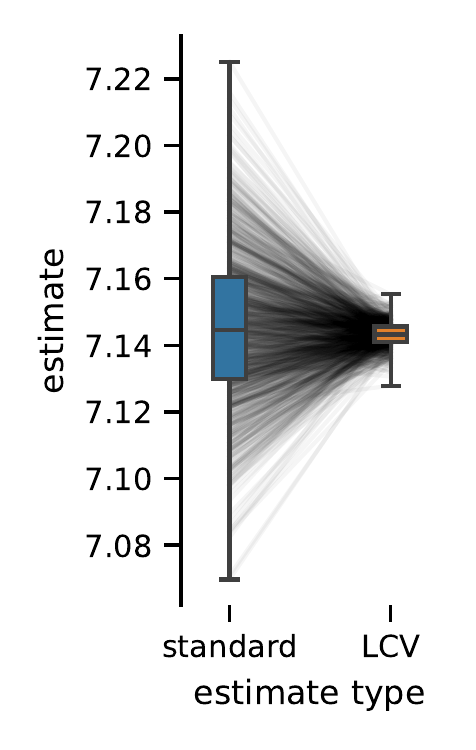}
    \caption{\new{(left) The correlation structure of control variates $Z_{0.4}$ and $Z_{2.7}$ with the objective random variable $M_2$ in Model~\ref{model:dm}.} (right) Estimates using the same trajectories with and without using the control variates.}
\end{figure}

\section{Finding Efficient Control Variates}\label{sec:algo}
\begin{figure}
    \centering
    \includegraphics[scale=0.65]{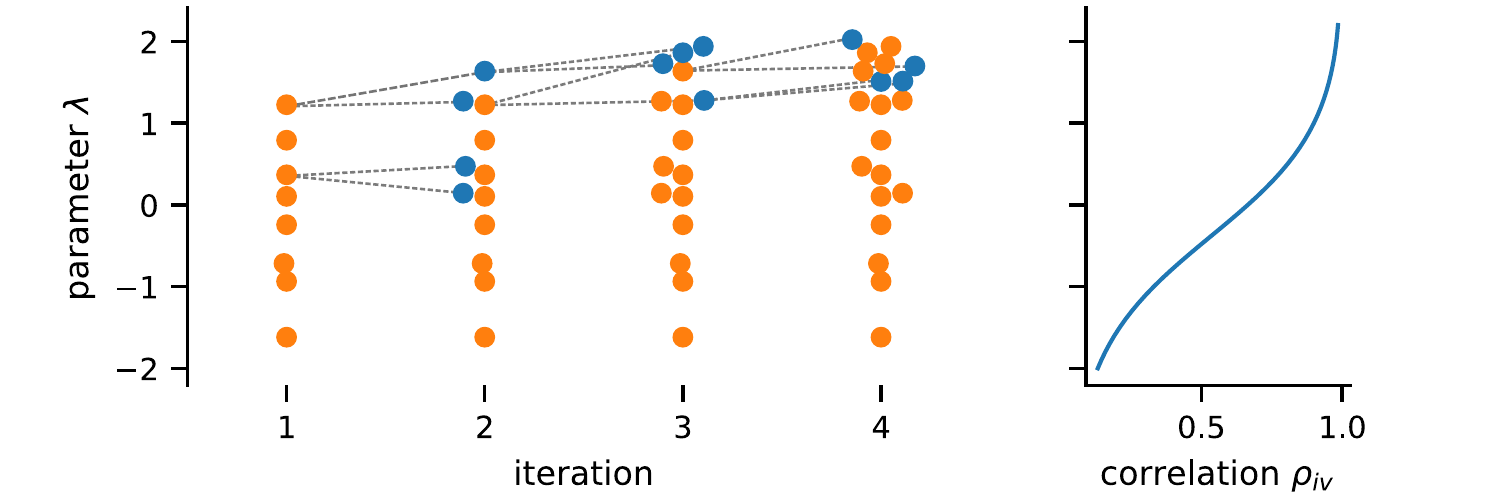}
    \caption{An illustration of the resampling procedure for the time-weighting parameter $\lambda$ using Model~\ref{model:dm}. Areas giving higher correlations are resampled through multiple rounds. The newly sampled values are given in blue. In each round only the new candidates are evaluated.}
    \label{fig:resampling}
\end{figure}

\new{In this work, we propose a novel algorithm to synthesize an efficient set of control variates.
As we have seen in the previous section, effective control variates have a high correlation
with the target random variable.
In the case of a single variate, the variance reduction is directly proportional to $1-\rho^2$, where
$\rho$ is the correlation.
In our case, infinitely many choices of $Z$ are available.
Our goal is to choose a subset that
satisfies two objectives:
Firstly, every selected control variate should \vw{significantly}\MB{Der Begriff ist immer so stark konnotiert...} reduce the estimator's variance.
Secondly, the subset should not be too large, i.e.  we want to avoid redundancies to achieve 
 a good  overall computational efficiency of the variance reduction.

Accordingly, we define an \emph{efficiency} value to estimate
whether the reduction in variance is compensating for the associated cost increase.
A natural baseline of any variance reduction is that it outweighs its associated additional costs.
Let $\sigma_Y^2$ be the variance of $Y$.
The \emph{efficiency} of the method is the ratio of the necessary
cost to achieve a similar reduction with the CV estimate $Y_{\text{CV}}$ compared to
the standard estimate $Y$~\cite{l1994efficiency,mcbook}, i.e.
\begin{equation}\label{eq:efficiency}
E=\frac{c_0\sigma_Y^2}{c_1\sigma^2_{Y_{\text{CV}}}}\,.
\end{equation}
The cost ratio $c_0/c_1$ depends on both the specific implementation and the technical setup.
The cost increase is  mainly due to the computation of the integrals in \eqref{eq:dis_int}.
The accumulation over the trajectory directly increases the cost of a single simulation,
which is the critical part of the estimation.

If the computation of a variate does not adequately compensate for its computation
with variance reduction, we do not want to include it.
Balancing both objectives is challenging because control variates often correlate with each other.
Such correlations expose redundancy between different variates.
This also becomes clear, when considering that the overall variance reduction  depends on
the coefficient of multiple correlation.

Here we follow a resampling paradigm:
We start by building up a set of candidates using a particle splitting approach.
After each splitting step, we generate a small number of SSA samples to estimate correlations.
Promising candidates are chosen based on the \emph{improvement} they provide and their time-weighting parameter $\lambda$ is resampled (see Fig.~\ref{fig:resampling}).
The main benefit of this bottom-up approach is its lower dependence on the initial sampling distribution of $\lambda$. \vw{compared to the approach in Zitat?}
Moreover, the procedure spends less time evaluating unpromising candidates.
After   generating a set of control variates, the overall covariance matrix is estimated using stochastic simulations.
Using this information, we construct an efficient subset using a greedy scheme, taking into account the redundancies between control variates.
We discuss Algorithm~\ref{alg:ssa:lcv} in more detail below.
}

\begin{algorithm}[ht]
\SetKwInOut{Input}{input}
\SetKwInOut{Output}{output}
\Input{$n, d, n_{\max}, n_{\lambda}, n_{c}, n_s, n_r$}
\Output{estimate using linear control variates}
  $L\leftarrow\{\lambda_i\sim \pi_\lambda \mid 1\leq i< n_{\lambda} \} \cup \{ 0 \}$\label{line:lambda_sample}\tcp*[r]{initialization}
  $P \leftarrow \{ ({m}, \lambda) | 1\leq\lvert {m} \rvert \leq n_{\max}, \lambda\in L\}$\label{line:init_covs}\;
  $P_{\text{all}} = \emptyset$\;
  \For{$i=1,\dots,n_r$\label{line:main_loop} \tcp*[r]{resampling}}{
    $(\hat{\Sigma},\hat{V}, \hat{Z}) \leftarrow$ \textsc{SSA}($\pi_0,T,P,d$)\;

     $P_{\text{all}}\leftarrow P_{\text{all}}\cup P$\;
     $I_{\text{cands}}\leftarrow \{ k\sim \hat\gamma_{kv}/ \sum_{\ell}\hat{\gamma}_{\ell v} \mid 1\leq k \leq |P_{\text{all}}|, j=1,\dots,n_c\}$\label{line:resample}\;
     $P\leftarrow \bigcup_{k\in I_{\text{cands}}}\bigcup_{l=1}^{n_s}
     \{(m_k, \lambda_k') \mid \lambda_k'\sim N(\lambda_k, 0.5)\}$\label{line:end_resampling}\;
  	}
    $(\hat{\Sigma},\hat{V}, \hat{Z}) \leftarrow$ \textsc{SSA}($\pi_0,T,P_{\text{all}},5d$)\label{line:search_loop}\tcp*[r]{covariance estimation}
     $P^*=\emptyset$\;
     \While{$\exists i:
        (m_i, \lambda_i)\in P_{\text{all}}\setminus P^*\land
        \hat{\gamma}_{iv} \prod_{j=1;(m_j, \lambda_j)\notin P^*}^{\left|P_{\text{all}}\right|}
        \hat{\gamma}_{ij}^{-1}
        >\epsilon$\tcp*[r]{selection}\label{line:selection_loop}}{
        $k \leftarrow \argmax_i \hat{\gamma}_{iv}
        \prod_{j=1;(m_j, \lambda_j)\notin P^*}^{\left|P_{\text{all}}\right|}
        \hat{\gamma}_{ij}^{-1}$\label{line:selection}\;
        $P^*\leftarrow P^* \cup
            \{(m_k,\lambda_k)\}$\;
    }
    $(\hat{\Sigma},\hat{V}, \hat{Z}) \leftarrow$ \textsc{SSA}($\pi_0,T,P^*,n$)\tcp*[r]{estimation}\label{line:sim}
    \Return $\hat{V} - {(\hat{\Sigma}_{{Z}}^{-1}\hat\Sigma_{{Z}V})}^{\top}\hat{{Z}}$\label{line:compute_lcv}
    \caption{\label{alg:ssa:lcv}Estimate the mean of species $i$ at time $T$}
\end{algorithm}

\paragraph{Initialization} A tuple $({m}_k, \lambda_k)$ of a moment vector ${m}_k$ and a time-weighting parameter $\lambda_k$
uniquely identifies a control variate $k$.
The algorithm starts out with an initial small set of control variates.
That is, we use $w(t)=e^{\lambda_kt}$ and $f(x)=x^{m_k}$ in \eqref{eq:poly_con}.
For a given set of time-weighting parameters $L$, we use all moments up to some fixed order $n_{\max}$ (line~\ref{line:init_covs}).
For a fixed moment vector $m_k$ the time-weighting parameter $\lambda_k$ can lead to vastly different correlations $\rho_{kv}$ with the quantity of interest.
The best choices of $\lambda$ are usually not known beforehand.
Therefore, we sample an initial set of $\lambda$'s
from a fixed distribution $\pi_{\lambda}$ (line~\ref{line:lambda_sample}).
\new{
Here, we use a standard normal distribution because its mean is the neutral weighting of $\lambda=0$ and extreme values are unlikely.
}

\new{
\paragraph{Resampling} Promising candidates are chosen from all control variates based on the estimated \emph{improvement ratio} \vw{re-use the term improvement ratio in Pseudocode?} they
provide, i.e.\ 
\begin{equation}
\hat{\gamma}_{kv} = ({1 - \hat{\rho}_{kv}^2})^{-1}
\end{equation}
following \eqref{eq:lcv_norm_varred}.
Specifically, control variate $k$ is chosen with probability proportional to
${\hat{\gamma}}_{kv}$ (line~\ref{line:resample}).
The covariances of (only) the new variates are roughly estimated using very few (e.g., $d= 10$) SSA samples. 
For the selected variates $I_{\text{cands}}$, the time-weighting parameter is resampled using a step distribution.
There is some freedom in the specifics of this resampling procedure.
In particular, the number of splits $n_c$ and descendants $n_s$ for each candidate control the number of additional candidates.
The algorithm performs $n_r$ rounds of resampling.
Figure~\ref{fig:resampling} illustrates this part of the algorithm.

\paragraph{Covariance Estimation} After sampling a set of candidates this way, we need to select the most promising ones.
For this, we are interested in covariances between all control variates, as well.
Since the resampling does not provide us with such estimates, we evaluate all candidates together for a fixed number of simulations (line~\ref{line:search_loop}).

\paragraph{Selection} The selection part of the algorithm (line~\ref{line:selection_loop}) proceeds in a greedy fashion wrt.\ the potential estimated improvement $\hat{\gamma}_{iv}$ given by any variate.
However, covariates often have high mutual correlations.
For example, $Z_{\lambda}$ and $Z_{\lambda+\epsilon}$ for a small $\epsilon$ are typically highly correlated --- often more with each other than with the objective.
We want to avoid this unnecessary computational overhead from computing nearly redundant information and numerical problems due to the covariance matrix inversion (see \eqref{eq:lcv_asym}).
As a solution, we normalize the estimated improvement vector $(\hat{\gamma}_{iv})_i$ by
the product of the fractions of explained variances by the already selected covariates.
Therefore we choose the most promising candidate given a selection $P^{*}$ as
\begin{equation}
\argmax_{1\leq i \leq \left|P_{\text{all}}\right|}
\hat{\gamma}_{iv}
    \prod_{\substack{1\leq j\leq {\left|P_{\text{all}}\right|}\\(m_j, \lambda_j)\notin P^*}}
    \hat{\gamma}_{ij}^{-1}
\end{equation}
in line~\ref{line:selection}.
This selection is done, until some lower threshold $\epsilon$ is reached (line~\ref{line:selection_loop}).

\paragraph{Estimation} Finally, we simulate the model $n$ times (line~\ref{line:sim}).
The resulting information enables an LCV estimation (line~\ref{line:compute_lcv}).
}

\section{Case Studies}\label{sec:study}
The simulation is implemented in the Rust programming language\footnote{\url{https://www.rust-lang.org}}.
The model description is parsed from a high level specification. 
Rate functions are compiled to stack programs for fast evaluation.
To estimate the base-line cost $c_0$, 1000 estimations were performed without
considering any control variates.

We first consider two case studies that have already been used in previous work \cite{backenkohler2019control}.
The first model is a simple dimerization process, albeit with an countably infinite state-space.
The second model is a switch model with a more complex bi-modal behavior.
We now describe the models and the estimated quantities in detail.
\begin{model}[Dimerization]
We first examine a simple dimerization model on an unbounded state-space
$$\varnothing\xrightarrow{10}\mathrm{M},\quad 2M\xrightarrow{0.1}\mathrm{D}$$
with initial condition $X_0^{\mathrm{(M)}}=0$.
\end{model}
Despite the models simplicity, the moment equations are not closed for this system
due to the second reaction which is non-linear.
Therefore a direct analysis of the expected value would require a closure.
For this model we will estimate $\expSym(X^{\mathrm{(M)}}_2)$.

The following models is bimodal, i.e.\ they each posses two stable regimes
among which they can switch stochastically.
We choose the initial conditions such that the process
will move towards either attracting region with equal probability.
\begin{model}[Distributive Modification]\label{model:dm}
This model was introduced in \cite{cardelli2012cell}.
It consists of the reactions
\begin{align*}
    \mathrm{X + Y} \xrightarrow{.001} \mathrm{B + Y}\,,\quad
    &\mathrm{B + Y }\xrightarrow{.001}\mathrm{ 2 Y}\,,\\
   \mathrm{Y + X }\xrightarrow{.001} \mathrm{B + X}\,,\quad
   &\mathrm{B + X }\xrightarrow{.001} \mathrm{2 X}
\end{align*}
with initial conditions $X^X_0=X^Y_0=X^{\mathrm{(B)}}_0=100$.
\end{model}
The expected value of interest here is $\expSym(X^{\mathrm{(X)}}_{50})$.

\begin{figure}[htb]
    \centering
    \includegraphics[scale=0.65]{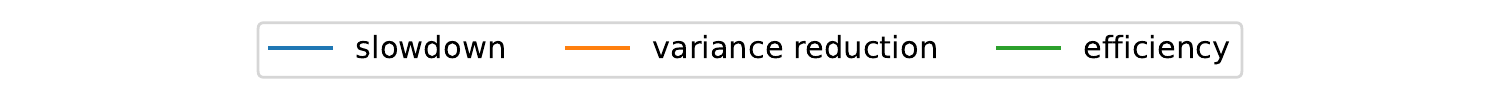}\\
    \includegraphics[scale=0.65]{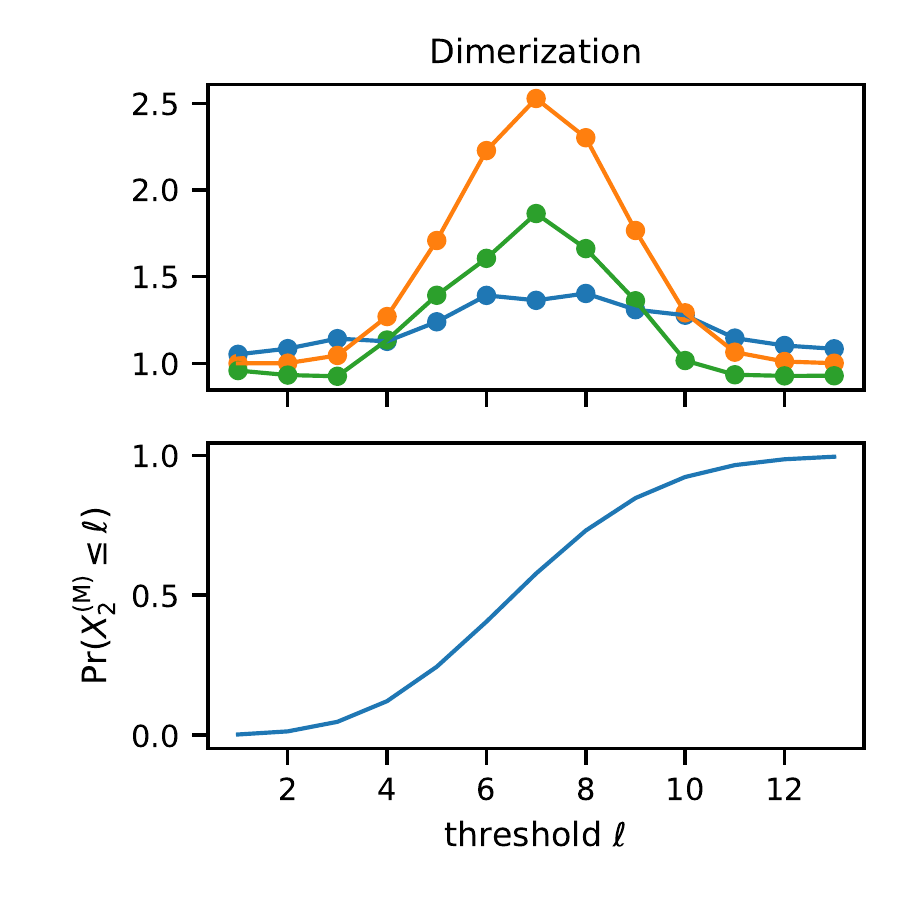}
    \includegraphics[scale=0.65]{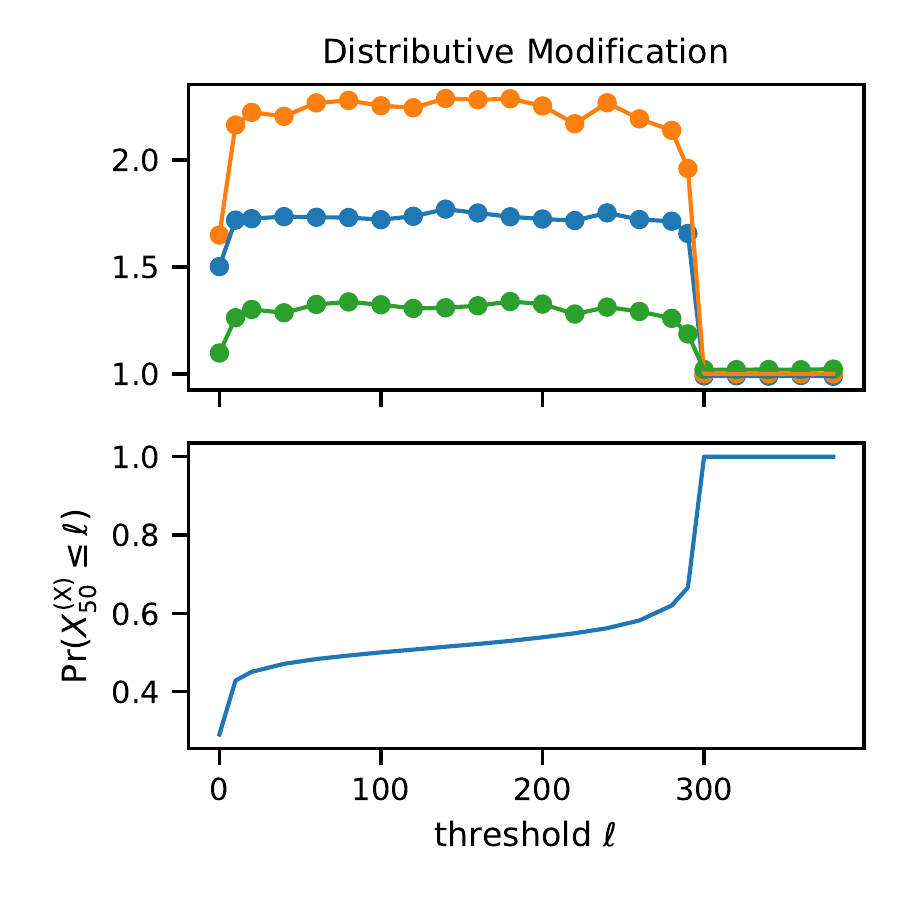}
    \caption{The methods efficiency for the estimation of threshold probabilities. For each threshold $\ell$ at least 200 estimations were performed.}
    \label{fig:thresholds}
\end{figure}

\new{

We applied the presented algorithm for an estimation using $n=\num{10000}$ simulations.
Initially $n_{\lambda}=10$ samples for the time-weighting parameter were drawn from a standard normal distribution ($\pi_{\lambda} = N(0,1)$).
Constraints corresponding to each first-order moment, i.e.\ the process' expectations were generated ($n_{\max}=1$).
The covariance estimation during resampling used $d=10$ samples.

We evaluated the algorithm both with and without resampling for these first two case studies.
The algorithm without resampling  leaves out lines~\ref{line:main_loop}--\ref{line:end_resampling} from Alg.~\ref{alg:ssa:lcv}.
The evaluation without resampling provides a good point of comparison to our previous heuristics performance on these cases.
In the case of dimerization we observe a variance reduction of $\approx 27.67$ compared to a best case reduction of $\approx 28.75$ in our previous work.
This close performance however has to balance very different slowdown factors: With our new heuristic the slowdown is a factor of $\approx 1.34$ while in the previous case it was $\approx 1.95$.
Therefore the new method clearly outperforms in terms of efficiency ($\approx 20.5$ (new) versus $\approx 14.86$ (old)).
This is mainly due to the higher number of covariates used by the simple threshold heuristic.
In contrast the new method takes into account redundancies between covariances while still retaining good performance.
This becomes apparent when comparing the average number of used variates ($\approx 3.34$ (old) versus $\approx 1.98$ (new)).

The variance reduction factor for the distributive modification model is similar at $\approx 2.63$ (old) versus $\approx 2.66$ (new).
Noticeably, the new method uses on average fewer CVs ($\approx 2.74$) than the previous heuristic with the best efficiency ($\approx 3.23$).
The overall efficiency of the new algorithm with $1.72$ is slightly lower than the previous best value of $1.77$, due to a higher slowdown.
It is however important to note, that the trade-off differs significantly between different  heuristics used in the previous algorithm.
Furthermore, the lower average number of control variates would reduce the slowdown factor further, if more trajectories are generated.

\begin{figure}[htb]
    \centering
    \includegraphics[scale=.65]{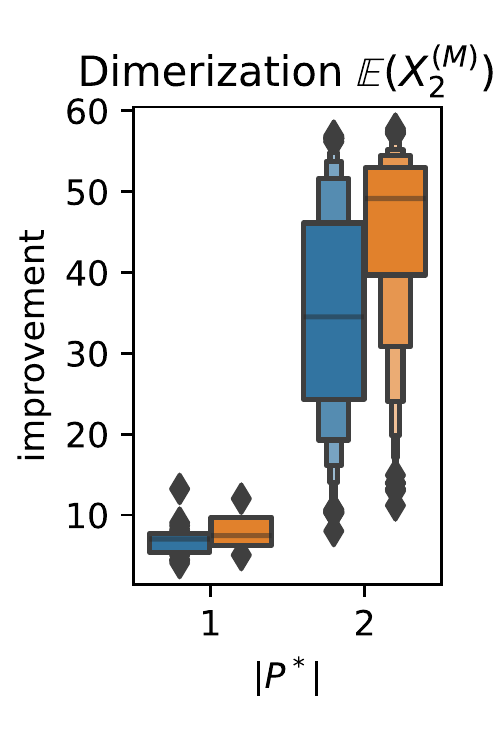}
    \includegraphics[scale=.65]{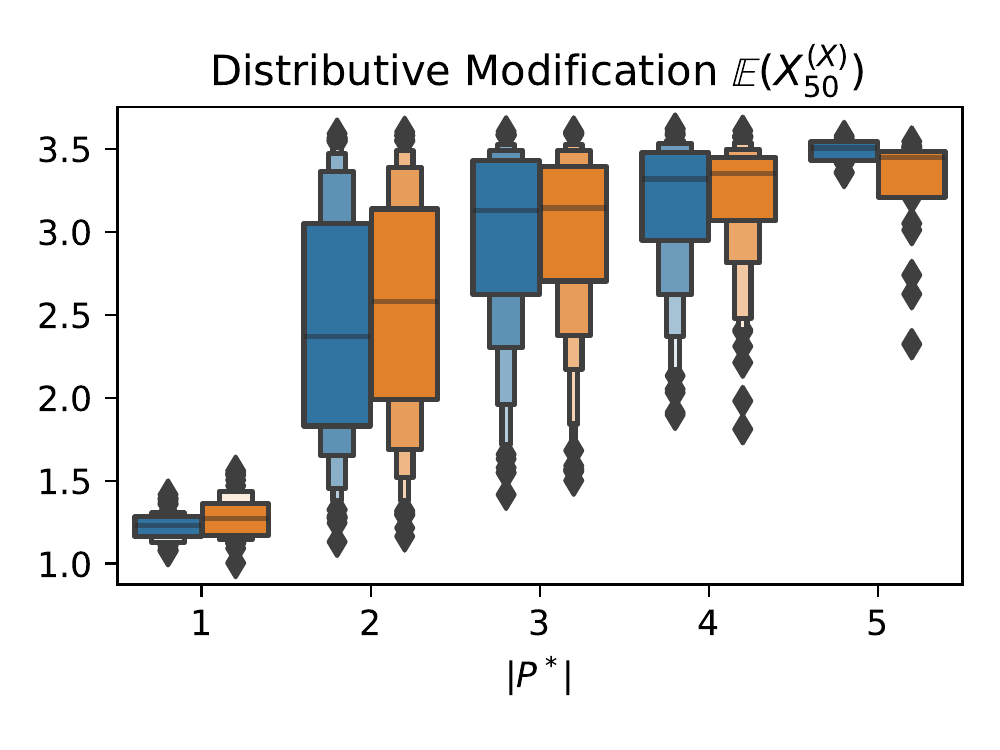}
    \includegraphics[scale=.65]{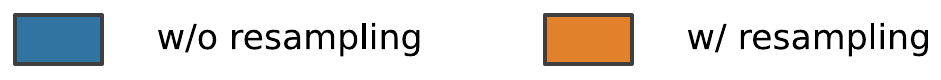}
    \caption{\label{fig:refinement}The variance reduction factor $\sigma_{Y}^2 / \sigma_{Y_{LCV}}^2$ over different numbers of selected covariates with and without the resampling procedure.}
\end{figure}

In Figure~\ref{fig:refinement} we contrast the variance improvement ratio with and without the resampling algorithm. For the dimerization model, we see a clear improvement of variance reduction.
This improvement is due to the fact that the strongest correlations are present for $\lambda \approx 2.5$ (cf.\ Figure~\ref{fig:resampling}).
This region of the time-weighting parameter space is less likely to be sampled by the initial
samples from the standard normal distribution.
Therefore the resampling procedure is especially beneficial if the better parameters $\lambda$ are farther from the origin.
In case of the distributive modification case study, we see a slight improvement.
Here, the best parameters $\lambda$ are close to zero and thereby more likely to be sampled by a standard normal.
Still, the resampling improves covariate performance for the most frequent cases of 2--4 covariates being selected (the case of 5 covariates has only a few instances).
Note, that the additional cost incurred by the resampling procedure is comparatively small, because
at most 4 candidates are evaluated in each iteration.

Next, we turn to the estimation of probabilities.
In particular, we consider the event of a species being below a threshold $\ell$ at time $t$ (species $M$ for the dimerization and $X$ for the distributive modification).
In Figure~\ref{fig:thresholds} we summarize the results of this study for varying levels $\ell$.
In both case studies we observe that control variates are efficient for probabilities not close to either zero or one.
In this case control variates are able to reduce the variance of the estimated probabilities whilst maintaining a beneficial reduction-slowdown trade-off.
This region is larger for the distributive modification model because of its bimodal behavior.
If the probability to be estimated is close to either one or zero, the event occurs too rarely or too often, respectively, to adequately explain variance using linear correlations.
We note, that the worst case efficiency is close to one.
This is due to the algorithm throwing out all covariate candidates leaving us with a standard estimation.
Only the initial covariate evaluation and resampling causes a slowdown, driving efficiency slightly below one.
Naturally this cost decreases with more samples $n$.

Control variates based on test functions restricted to intervals \vw{vielleicht nochmal auf die Papiere verweisen}\MB{welche?} did not lead to an improvement (data not shown).

}

\new{

Finally, with the lac operon model we consider a larger case study. This model consists of 11 species and 25 partly non-linear reactions.
\begin{model}[Lac operon]This is a well-known model of genetic regulation with positive feedback \cite{stamatakis2009comparison}. Its reactions are
 \begin{gather*}
 \varnothing \xrightleftharpoons[k_19]{k_1} \mathrm{M_R}\,, \quad
 \mathrm{M_R} \xrightarrow{k_2} \mathrm{M_R + R}\,,\quad
 2 \mathrm{R} \xrightleftharpoons[k_4]{k_3} \mathrm{R_2}\,,\quad
 \mathrm{R_2 + O} \xrightleftharpoons[k_6]{k_5}\mathrm{ R_2O}\,,\\
 \mathrm{2 I + R_2} \xrightleftharpoons[k_8]{k_7} \mathrm{I_2R_2}\,,\quad
\mathrm{2I +R_2O} \xrightleftharpoons[k_{10}]{k_9} \mathrm{I_2R_2 + O}\,,\quad
\mathrm{O}\xrightarrow{k_{11}} \mathrm{O + M_Y}\,,\\
\mathrm{M_Y} \xrightarrow{k_{13}}\mathrm{M_Y + Y}\,,\quad
\mathrm{Y + I_{{ex}}}\xrightleftharpoons[k_{15}]{k_{14}}\mathrm{YI_{{ex}}}\,,\quad
\mathrm{YI_{{ex}}}\xrightarrow{k_{16}}\mathrm{Y+I}\,,\quad
\mathrm{I_{{ex}}}\xrightleftharpoons[k_{18}]{k_{17}}\mathrm{I}\,,\\
\mathrm{M_Y}\xrightarrow{k_{20}}\varnothing\,,\quad
\mathrm{R}\xrightarrow{k_{21}}\varnothing\,,\quad
\mathrm{R_2}\xrightarrow{k_{22}}\varnothing\,,\quad
\mathrm{Y}\xrightarrow{k_{23}}\varnothing\,,\\
\mathrm{YI_{ex}}\xrightarrow{k_{24}}\mathrm{I}\,,\quad
\mathrm{I_2R_2}\xrightarrow{k_{25}}\mathrm{2I}\,,\quad
\mathrm{R_2O}\xrightarrow{k_{12}}\mathrm{R_2O+M_Y}\,.
\end{gather*}
Initially, $X^{(\mathrm{O})}_0=1$ and $X^{(\mathrm{I_{{ex}}})}=\num{48177}$ while all other abundancies are zero.
The parameters are $
  k_1  = 0.111$,
 $ k_2  = 15.0$,
  $k_3  = 103.8$,
  $k_4  = \num{0.001}$,
  $k_5  = 1992.7$,
  $k_6  = 2.4$,
  $k_7 =k_9  = \num{1.293d-7}$
  $k_8  = 12$,
  $k_{10} = 9963.2$,
  $k_{11} = 0.5$,
  $k_{12} = 0.01$,
  $k_{13} = 30.0$,
  $k_{14} = 0.249$,
  $k_{15} = 0.1$,
  $k_{16} = \num{6.0d4}$,
  $k_{17} =k_{18} = 0.92$,
  $k_{19} = k_{20} = 0.462$,
  $k_{21}=k_{22}= k_{23} = k_{24} = k_{25} = 0.2$.
\end{model}
We estimate the abundancy of LacY after one time unit, i.e.\ $\expSym(X^{(Y)}_1)$.
It is encoded by Y and facilitates
the lactose import via reactions 14 and 16.
A typical simulation of the system up to time-horizon $T=1$ takes well above one minute of computational time.
Therefore we reduce the number of used trajectories to $n=1000$.
The other settings remain as above.

Despite the high dimensionality, we observe a good efficiency value of $E\approx 4.85$.
The slowdown caused by the method is approximately 1.98.
A big part of this slowdown is due to the initial search of covariates.
Initially 10 covariates are generated for each first order moment, i.e.\ each of the
11 species. The number of additionally resampled covariates is similar to previous case studies.
Thus the main cost of the initial resampling and selection is due to the first iteration of the
resampling loop and the simulation loop of the selection procedure.
This part naturally has still potential for optimization:
Not all known covariates need to be reconsidered at the selection stage.
Instead, unpromising candidates could be discarded prior to that stage.

Still, the high variance reduction by a factor of approx.\ ${9.64}^{-1}$ more than compensates for this increase in computational cost, leading to the good overall efficiency.
This shows that, even for more complex models, the method is applicable and can extremely beneficial for Monte Carlo estimation.

}

\section{Conclusion}\label{sec:conclusion}
\new{In the context of Monte Carlo simulation, variance reduction
techniques offer an elegant way of improving the performance without
introducing approximation errors in addition to the statistical uncertainty. 

For stochastic reaction networks, we show that it is possible to exploit constraints derived from the statistical moment equations for the construction of control variates.
We propose a robust method to select an appropriate subset from the large set of all possible variates.
In particular, we improve an initial subset by selecting particularly effective variates and removing redundant variates.
By resampling the time-weighting parameter $\lambda$ we ensure that appropriate values  are 
flexibly explored.
In the worst case, all variates are dropped and the performance approaches the standard SSA.
In most cases, however, a suitable subset is found together with the corresponding choices of $\lambda$.

We  analyze the performance of the method when estimating event probabilities and not only average molecule counts. Our largest case study has 11 species and 
24 reactions.

In the future, we will further explore the algorithmic design space.
For example, the resampling distribution could be adjusted using decaying standard deviations.
Furthermore, we will look at different test functions weighting the state space more flexibly.

}
\subsubsection{Acknowledgements} This work was supported by the DFG project MULTIMODE.

%
%

\bibliographystyle{splncs04}
\bibliography{paper.bib}
\end{document}